# Raman Modes Non-classicality through Entangled Photons Coupling to Plasmonic Modes

Ahmad SalmanOgli[1,2]
[1]Faculty of engineering, Çankaya University Electrical and Electronics engineering department, Ankara, Turkey
[2]Chemical Engineering Department, Hacettepe University, 06800, Ankara, Turkey.

**Abstract**
In this article, non-classical properties of Raman modes are investigated. The original goal, actually, is to identify how and by which method we can induce non-classicality in Raman modes. We introduce a plasmonic system in which Raman dye molecules are buried between two shells of the plasmonic materials similar to onion-like core/shell nanoparticle. This system is excited by the entangled photons, followed by analyzing its dynamics of motion using the Heisenberg-Langevin equations by which the time evolution of the signal-idler mode and Raman modes are derived. Interestingly, the entangled photons are coupled to the plasmonic modes which are used to improve the non-classicality. It is shown that the exciting system with the entangled photons lead to inducing the non-classicality in Raman modes and entanglement between them. This behavior is attributed to the non-classicality of input modes that is coupled to the Raman modes considering the correlation between the incident wave frequency and Raman modes frequency. Notably, these quantum properties are dramatically affected by the environment temperature and Raman molecules location around the plasmonic nanoparticles. Modelling results demonstrate that a temperature increase has a drastic effect on system dynamics. Moreover, it is found that the entanglement between modes in system surely is affected by the coupling between the incident modes and plasmonic modes generated by the core/shell nanoparticles. Finally, as an important result, it is revealed that the Raman modes such as stoke and anti-stoke modes show a revival behavior, which is a quantum phenomenon.

## Introduction

Raman spectroscopy can be simply defined as an interaction of light with matter and resonance of molecular bonding, which can lead to the Raman scattering modes such as Stoke, Anti-stoke, and Phonon modes. Considering that Raman scattering without any signal amplification is an extremely inefficient process due to its small cross-sections, it is necessary to enhance the efficiency of Raman scattering. This amplification, typically, may be made through the plasmonic nanoparticle (NPs) effect [1-4]. Through this process, Raman molecules typically attach to the NPs surface and, because of their localized high-intensity near-field, amplify the Raman reporter scattering similar to the amplification of radio frequency signals by antenna [6, 7]. In the present work, the Raman dye molecules are embedded in the interior section of the core/shell NPs such as Au/air/Au. The quantum properties of the plasmon-plasmon interaction and their effect on the dye molecules as NPs-dye coupling factor were previously investigated in our recent work [8]. To entangle Raman modes, as the original aim of the present research, we can reasonably apply the entangled input whereby the incident wave non-classicality can couple to Raman modes. Indeed, the coupling quality and quantity dramatically depend on system parameters, which should be designed effectively. Accordingly, we decided to excite the designed system with the entangled photons (signal-idler) wave [9-13] rather than a traditional laser wave. It has been reported that the entangled two-photon system, because of its unique quantum properties, has an indispensable effect on different applications such as quantum communications [14, 15], biology [16], and quantum imaging [10, 17]. Originally, quantum entanglement is a physical phenomenon that occurs when pairs or groups of particles are generated or interacting such that the quantum state of each particle cannot be described independently; rather, the definition can be done as a whole. Quantum entanglement [18-20] refers to the non-classic and non-local strong correlations among quantum systems. With knowledge about these unique quantum properties of the entangled two-photon, we try to excite the considered system with this type of non-classical signal wave, which is initially coupled to the plasmonic modes, followed by coupling the plasmonic entangled mode to the Raman molecules. We assume that through this approach the Raman signals will be enhanced because of the entangled incident wave and its coupling to the plasmonic modes. Therefore, we have to study the Raman scattering quantum picture [21, 22]. Additionally, it should note that using entangled incidence leads to a good signal-to-noise ratio [23], which is an important factor in Raman spectroscopy. Hence, with knowledge of this information, we want to study the Raman modes entanglement, and in this work, it will be focused on the quantum properties of the Raman modes and their non-classical properties that lead to the entanglement of the contributed modes.

**Theory and Background**
*Model description*
In this section, the theoretical and system backgrounds such as model description and system dynamics are presented. The system schematic is illustrated in Fig. 1a. Firstly, we suppose that based on the common way we can produce an entangled two-photon system through the interaction of a high power laser incident on a nonlinear material. In the next step, we show the interaction of the entangled photon's wave separated by a beam splitter using the core/shells NPs containing the Raman dye molecules. We embed Raman molecules in the middle gap of NPs because I) The maximum plasmon resonance can be coupled to dye molecules in this region [6, 7] and (II) The reproducibility factor is very important in Raman scattering mechanism, which is easily accessible by this structure [7]. In addition, in terms of core/shells NPs, we can easily manipulate the plasmon-plasmon interaction by which the plasmonic resonance peaks will be controlled and fitted with the signal and idler waves for maximum plasmon coupling. In this study, the Raman molecule is considered as a three-level energy in which $E_2-E_1-\hbar\omega_{sp} = E_2-E_3-\hbar\omega_{ip} = \hbar\Delta$. Also, it assumed that $\omega_{sp} > \omega_{ip}$ and $\hbar\Delta \gg E_3-E_1$. Based on this assumption, Level 2 is far off resonance and will be adiabatically removed. In this relation, $\Delta$ is the detuning parameter. Therefore, in the adiabatic condition, the presented Raman molecule is considered as a two-level energy molecule. As can be seen from Fig. 1b, the plasmon resonance peaks can be engineered by changing the gap and outer shell thicknesses. On the other hand, one can easily design the NPs morphology to match the plasmon resonance with the frequency of the entangled photon (signal and idler) to transfer the maximum amount of field to dye molecules. The methodology applied in the present work includes: I) Generation of entangled photons, II) Separation of the entangled photons by a polarization dependent beam splitter, III) Coupling the entangled photons with plasmonic modes through core/shell NPs, and IV) Exciting the Raman molecules by the plasmonic-entangled modes.

*System dynamics*
Raman scattering quantum picture is fully described in [21, 22]. Briefly, an incident photon on the Raman molecules is annihilated, leading to a phonon with frequency $\omega_{ph}$ and a Stoke photon with frequency $\omega_s = \omega_{sp\ or\ ip} - \omega_{ph}$. Moreover, an incident photon with generated phonon may be annihilated; and create an Anti-stoke photon with frequency $\omega_{as} = \omega_{sp\ or\ ip} + \omega_{ph}$. Based on this knowledge and regarding the schematic shown in Fig. 1a, we can derive the Hamiltonian that efficiently describes this system as:

$$H_0 = \hbar\omega_{sp} a_s^+ a_s + \hbar\omega_{ip} a_i^+ a_i + \hbar\omega_s b^+ b + \hbar\omega_{as} c^+ c + \hbar\omega_{ph} d^+ d + \hbar\omega_0 \sigma_0/2$$

$$H_{f\_f} = \hbar\kappa_s(a_s b^+ d^+) + \hbar\kappa_i(a_i b^+ d^+) + \hbar\kappa_{as}(a_s d c^+) + \hbar\kappa_{ai}(a_i d c^+) + H.c.$$

$$H_{f\_a} = (-2\hbar g_s g_i/\Delta)(\sigma_+ a_s a_i^+ + \sigma_- a_i a_s^+)$$

$$H_{res} = \int \hbar\omega'(F^+ F) d\omega', \quad H_{sr} = i\int \hbar\omega'(T_1 F^+ a_s + T_2 F^+ a_i + T_3 F^+ \sigma_- + T_4 F^+ b + T_5 F^+ c + T_6 F^+ d) d\omega' + H.c.$$

$$H_{drive} = i\hbar[\sqrt{\kappa}(a_s^+ \varepsilon e^{-i\omega_{sL}t} - a_s \varepsilon^* e^{i\omega_{sL}t}) + \sqrt{\kappa}(a_i^+ \varepsilon e^{-i\omega_{iL}t} - a_i \varepsilon^* e^{i\omega_{iL}t})]$$

$$H_E = \hbar\chi^{(2)} E_f a_s^+ a_i^+, \quad E_f = (|E_p|/2).\sqrt{\omega_{sp}.\omega_{ip}}$$

(1)

where $H_0$, $H_{f-f}$, $H_{f-a}$, $H_{res}$, $H_{drive}$, $H_{sr}$, and $H_E$ are the free space Hamiltonian, field-field interaction Hamiltonian, field-molecule interaction Hamiltonian, reservoir mode Hamiltonian, external field system interaction Hamiltonian, system-reservoir mode interaction Hamiltonian, and second-order nonlinear interaction Hamiltonian, respectively. In the case of latter Hamiltonian, optical parametric down conversion (OPDC) can be considered as a powerful method for generation of the entangled two-photon [9, 11]. The pump power is intense enough, so nonlinear effect leads to creation of the pairs of correlated photons. Many works have proved theoretically and experimentally that the pairs of photons (signal and idler) are entangled, and thus show the non-classical properties [11]. In this equation, $\chi^{(2)}$ indicates the second-order susceptibility; [$\kappa_s$, $\kappa_i$] and [$\kappa_{as}$, $\kappa_{ai}$] denote the coupling constants for the Stoke and Anti-stoke processes for signal and idler exciting; $T_i$ (i=1-6) is the coupling strength between system's mode and reservoir model; in this equation one can see $\sigma_0 = \sigma_{33} - \sigma_{11}$, $\omega_0 = \omega_{sp} - \omega_{ip}$, and $g_i$ (i = s or i) are the NPs-molecule coupling strength for different signals induced by the electric-dipole transition matrix [8]; and $\sigma_{ij}$ is raising and lowering operator between the defined levels. Also, $\kappa$ is the total decay rate of the medium due to the Ohmic losses and scattering into free space, which is derived from the reservoir modes coupling to NPs. Finally, $a_s$, $a_i$, b, c, d, and F are a notation for signal, idler, Stoke, Anti-stoke, phonon, and reservoir modes, respectively. It is clear that the signal and idler modes are coupled to the plasmonic modes which are schematically illustrated in Fig. 1b. Obviously, these operators obey the Bosonic commutation relation, for example, [c, c$^+$] = 1. It should be noted that damping rate for the Stoke, Anti-stoke, and phonon modes are very small and negligible [21]. By neglecting the fast oscillating terms in Eq. 1 at $\pm \omega_{sL}$, $\pm \omega_{iL}$ the system Hamiltonian becomes:

$$H_0 = \hbar \Delta_{sp} a_s^+ a_s + \hbar \Delta_{ip} a_i^+ a_i + \hbar \Delta_s b^+ b + \hbar \Delta_{as} c^+ c + \hbar \Delta_{ph} d^+ d + \hbar \Delta_0 \sigma_0 / 2$$

$$H_{f\_f} = \hbar \kappa_s (a_s b^+ d^+) + \hbar \kappa_i (a_i b^+ d^+) + \hbar \kappa_{as} (a_s d c^+) + \hbar \kappa_{ai} (a_i d c^+) + H.c.$$

$$H_{f\_a} = -2\hbar g_s g_i / \Delta (\sigma_+ a_s a_i^+ + \sigma_- a_i a_s^+)$$

$$H_{res} = \int \hbar \omega' (F^+ F) d\omega', \quad H_{sr} = i \int \hbar \omega' (T_1 F^+ a_s + T_2 F^+ a_i + T_3 F^+ \sigma_- + T_4 F^+ b + T_5 F^+ c + T_6 F^+ d) d\omega' + H.c.$$

$$H_{drive} = i\hbar [E_{cs}(a_s^+ - a_s) + E_{ci}(a_i^+ - a_i)], \quad E_{c(sori)} = \sqrt{2 P_c \kappa / \hbar \omega_{(sori)}}$$

(2)

where for the sake of simplicity we supposed that $\Delta_{sp} = \omega_{sp} - \omega_{sL}$, $\Delta_{ip} = \omega_{ip} - \omega_{iL}$, $\Delta_s = \omega_s - \omega_p$, $\Delta_{as} = \omega_{as} - \omega_p$, $\Delta_{ph} = \omega_{ph} - \omega_{sp}$, and $\Delta_0 = \omega_0 - \omega_{sL}$. $P_c$ is the input laser power and $E_{cs}$ and $E_{ci}$ are the input driving amplitude for signal and idler modes, respectively. To investigate the system's dynamics and non-classical properties, we derive the Heisenberg-Langevin equation of motions (using $i\hbar \dot{c} = [c, H]$, where c is a field operator) for the field operators, which gives:

$$\dot{a}_s = -i\Delta_{sp} a_s - i\kappa_s(bd) - i\kappa_{as}(d^+ c) - iG\sigma_- a_i - i\chi^{(2)} E_f a_i^+ + \int \omega T_1^* F d\omega + \sqrt{\kappa} \varepsilon_{in} + E_{cs}$$

$$\dot{a}_i = -i\Delta_i a_i - i\kappa_i(bd) - i\kappa_{ai}(d^+ c) - iG\sigma_+ a_s - i\chi^{(2)} E_f a_s^+ + \int \omega T_2^* F d\omega + \sqrt{\kappa} \varepsilon_{in} + E_{ci}$$

$$\dot{b} = -i\Delta_b b - i\kappa_s(a_s d^+) - i\kappa_i(a_i d^+) + \int \omega T_4^* F d\omega + \zeta_b$$

$$\dot{c} = -i\Delta_c c - i\kappa_{as}(a_s d) - i\kappa_{ai}(a_i d) + \int \omega T_5^* F d\omega + \zeta_c$$

(3)

$$\dot{d} = -i\Delta_{ph} d - i\kappa_s(a_s b^+) - i\kappa_{as}(a_s^+ c) - i\kappa_i(a_i b^+) - i\kappa_{ai}(a_i^+ c) + \int \omega T_6^* F d\omega + \zeta_d$$

$$\dot{\sigma}_- = -i\Delta_0 \sigma_- - iG\sigma_0 a_s a_i^+ + \int \omega T_3^* F d\omega$$

$$\dot{\sigma}_0 = -2iG(\sigma_+ a_s a_i^+ - \sigma_- a_i a_s^+) + 2\int \omega (T_3^* F \sigma_+ - T_3 F^* \sigma_-) d\omega$$

$$\dot{F} = -i\omega F(t) + T_1 a_s + T_2 a_i + T_3 \sigma_- + T_4 b + T_5 c + T_6 d, \quad F(t) = F(0) e^{(-j\omega t)} + \int_0^t (T_1 a_s + T_2 a_i + T_3 \sigma_- + T_4 b + T_5 c + T_6 d) e^{(-j\omega(t-t'))} dt'$$

In this equation, we consider the damping and noise process (optical noise and Raman modes noises) by which the dynamics of modes are affected because each mode interacts with its own environment. These modes in this equation are denoted by $\varepsilon_{in}$, $\zeta_b$, $\zeta_c$, and $\zeta_d$. By substituting F(t) from the last equation in Eq. 3 in the integral parts of above equations, the simple version of the system's equation of motions are presented as:

$$\begin{aligned}
\dot{a}_s &= -(i\Delta_{sp}+\kappa)a_s - i\kappa_s(bd) - i\kappa_{as}(d^+c) - iG\sigma_- a_i - i\chi^{(2)}E_f a_i^+ + \sqrt{\kappa}\varepsilon_{in} + E_{cs} \\
\dot{a}_i &= -(i\Delta_{ip}+\kappa)a_i - i\kappa_i(bd) - i\kappa_{ai}(d^+c) - iG\sigma_+ a_s - i\chi^{(2)}E_f a_s^+ + \sqrt{\kappa}\varepsilon_{in} + E_{ci} \\
\dot{b} &= -(i\Delta_b+\gamma_s)b - i\kappa_s(a_s d^+) - i\kappa_i(a_i d^+) + \zeta_b \\
\dot{c} &= -(i\Delta_c+\gamma_{as})c - i\kappa_{as}(a_s d) - i\kappa_{ai}(a_i d) + \zeta_c \\
\dot{d} &= -(i\Delta_{ph}+\gamma_{ph})d - i\kappa_s(a_s b^+) - i\kappa_{as}(a_s^+ c) - i\kappa_i(a_i b^+) - i\kappa_{ai}(a_i^+ c) + \zeta_d \\
\dot{\sigma}_- &= -(i\Delta_0+\gamma_z)\sigma_- - iG\sigma_0 a_s a_i^+ \\
\dot{\sigma}_0 &= -\gamma_z(\sigma_0+1) + 2iG(\sigma_- a_i a_s^+ - \sigma_+ a_s a_i^+)
\end{aligned} \quad (4)$$

where $\gamma_s$, $\gamma_{as}$, $\gamma_{ph}$, and $\gamma_z$ are the decay rate of the Stoke mode, Anti-stoke mode, phonon mode, and raising transition decay rate, respectively. As can be seen, Eq. 4 is a nonlinear equation. To linearize this equation, one can use the fluctuations associated with the field modes such as $c = <c> + \delta c$, where $<c>$ stands for field average in the steady-state condition and $\delta c$ indicates the fluctuation of the considered mode [24]. We can justify the linearized treatment due to following reasons. (i) We do not linearize the field itself. We put a $<c>$ and solve the system exactly (numerically). (ii) Afterwards, we linearize the fluctuations that are noise, (not the system itself!) around the $<c>$. Linearization of the noise in the system can be performed whenever the nonlinearities altering the noise are small. (iii) Linearization of the noise, since it is small in coherent states, is a well-established procedure applicable to systems with small noise and nonlinearities. So, in this work, we can use the quantum fluctuation approach to linearize the equation; it is because the plasmonic field created in the NP's gap is so intense [7], which indicates that the modes alteration can be oscillated as a small signal around the high intensity plasmon-plasmon field modes. Therefore, we can obtain the equations of motion based on the fluctuation in the form of linearization approximation as:

$$\begin{aligned}
\dot{\delta a}_s &= -(i\Delta_{sp}+\kappa)\delta a_s - i\kappa_s(B\delta d + D\delta b) - i\kappa_{as}(D^*\delta c + C\delta d^+) - iG\rho_{31}\delta a_i - i\chi^{(2)}E_f \delta a_i^+ + \sqrt{\kappa}\varepsilon_{in} \\
\dot{\delta a}_i &= -(i\Delta_{sp}+\kappa)\delta a_i - i\kappa_i(B\delta d + D\delta b) - i\kappa_{ai}(D^*\delta c + C\delta d^+) - iG\rho_{13}\delta a_s - i\chi^{(2)}E_f \delta a_s^+ + \sqrt{\kappa}\varepsilon_{in} \\
\dot{\delta b} &= -(i\Delta_b+\gamma_s)\delta b - i\kappa_s(\alpha_s\delta d^+ + D^*\delta a_s) - i\kappa_i(\alpha_i\delta d^+ + D^*\delta a_i) + \zeta_b \\
\dot{\delta c} &= -(i\Delta_c+\gamma_{as})\delta c - i\kappa_{as}(\alpha_s\delta d + D\delta a_s) - i\kappa_{ai}(\alpha_i\delta d + D\delta a_i) + \zeta_c \\
\dot{\delta d} &= -(i\Delta_{ph}+\gamma_{ph})\delta d - i\kappa_s(\alpha_s\delta b^+ + B^*\delta a_s) - i\kappa_{as}(\alpha_s^*\delta c + C\delta a_s^+) - i\kappa_i(\alpha_i\delta b^+ + B^*\delta a_i) - i\kappa_{ai}(\alpha_i^*\delta c + C\delta a_i^+) + \zeta_d
\end{aligned} \quad (5)$$

In this equation, $\alpha_s = <a_s>$, $\alpha_i = <a_i>$, $B = <b>$, $C = <c>$, $D = <d>$, $\rho_{31} = <\sigma_->$, $\rho_{13} = <\sigma_+>$ which are the contributed modes average, and these parameters will be calculated through the steady state equations. The steady state equations are introduced as:

$$\begin{aligned}
&-(i\Delta_{sp}+\kappa)\alpha_s - i\kappa_s(BD) - i\kappa_{as}(D^*C) - iG\rho_{31}\alpha_i - i\chi^{(2)}E_f\alpha_i^+ + E_{cs} = 0 \\
&-(i\Delta_{sp}+\kappa)\alpha_i - i\kappa_i(BD) - i\kappa_{ai}(D^*C) - iG\rho_{13}\alpha_s - i\chi^{(2)}E_f\alpha_s^+ + E_{ci} = 0 \\
&-(i\Delta_b+\gamma_s)B - i\kappa_s(\alpha_s D^*) - i\kappa_i(\alpha_i D^*) = 0 \\
&-(i\Delta_c+\gamma_{as})C - i\kappa_{as}(\alpha_s D) - i\kappa_{ai}(\alpha_i D) = 0 \\
&-(i\Delta_{ph}+\gamma_{ph})D - i\kappa_s(\alpha_s B^*) - i\kappa_{as}(\alpha_s^* C) - i\kappa_i(\alpha_i B^*) - i\kappa_{ai}(\alpha_i^* C) = 0 \\
&-(i\Delta_0+\gamma_z)\rho_{31} - iG\alpha_s\alpha_i^+ \rho_z = 0 \\
&-(\rho_z+1)\gamma_z + 2iG(\alpha_s^+\alpha_i\rho_{31} - \alpha_i^+\alpha_s\rho_{31}) = 0
\end{aligned}$$

(6)

Clearly, Eq. 5 is not linear and thus needs to be solved numerically. To solve this equation, we should initially solve the steady-state equations, and find $\alpha_s$, $\alpha_i$, B, C, D, and $\rho_{31}$ to substitute in Eq. 5. In the following, to analyze the system quadrature fluctuation and anti-bunching effect, we need to calculate the Hermitian conjugate of the field operators; one can derive these easily by taking the Hermitian of each field operator in Eq. 5. Therefore, all fields and their Hermitian conjugate operators can be presented as the following matrix (Eq. 7):

$$\begin{bmatrix} \dot{\delta a_s} \\ \dot{\delta a_s}^+ \\ \dot{\delta a_i} \\ \dot{\delta a_i}^+ \\ \dot{\delta b} \\ \dot{\delta b}^+ \\ \dot{\delta c} \\ \dot{\delta c}^+ \\ \dot{\delta d} \\ \dot{\delta d}^+ \end{bmatrix} = \underbrace{\begin{bmatrix} -\Gamma_s & 0 & -iG\rho_{31} & -i\chi^{(2)}E_f & -i\kappa_s D & 0 & -i\kappa_{as}D^* & 0 & -i\kappa_s B & -i\kappa_{as}C \\ 0 & -\Gamma_s^* & i\chi^{(2)}E_f^* & iG\rho_{13} & 0 & i\kappa_s D^* & 0 & i\kappa_{as}D & i\kappa_{as}C^* & i\kappa_s B^* \\ -iG\rho_{31} & 0 & -\Gamma_i & 0 & -i\kappa_i D & 0 & -i\kappa_{ai}D^* & 0 & -i\kappa_i B & -i\kappa_{ai}C \\ 0 & iG\rho_{13} & 0 & -\Gamma_i^* & 0 & i\kappa_i D^* & 0 & i\kappa_{ai}D & i\kappa_{ai}C^* & i\kappa_i B^* \\ -i\kappa_s D^* & 0 & -i\kappa_i D^* & 0 & -\Gamma_b & 0 & 0 & 0 & 0 & \Pi_1 \\ 0 & i\kappa_s D & 0 & i\kappa_i D & 0 & -\Gamma_b^* & 0 & 0 & \Pi_1^* & 0 \\ -i\kappa_{as}D & 0 & -i\kappa_{ai}D & 0 & 0 & 0 & -\Gamma_c & 0 & \Pi_2 & 0 \\ 0 & i\kappa_{as}D^* & 0 & i\kappa_{ai}D^* & 0 & 0 & 0 & -\Gamma_c^* & 0 & \Pi_2^* \\ -i\kappa_s B & -i\kappa_{as}C & -i\kappa_i B^* & -i\kappa_{ai}C & 0 & \Pi_1 & \Pi_2 & 0 & -\Gamma_{ph} & 0 \\ i\kappa_{as}C & i\kappa_s B & i\kappa_{ai}C^* & i\kappa_i B & \Pi_1^* & 0 & 0 & \Pi_2^* & 0 & -\Gamma_{ph}^* \end{bmatrix}}_{A_{i,j}} \times \underbrace{\begin{bmatrix} \delta a_s \\ \delta a_s^+ \\ \delta a_i \\ \delta a_i^+ \\ \delta b \\ \delta b^+ \\ \delta c \\ \delta c^+ \\ \delta d \\ \delta d^+ \end{bmatrix}}_{u(t)} + \underbrace{\begin{bmatrix} \sqrt{\kappa}\varepsilon_{in} \\ \sqrt{\kappa}\varepsilon_{in}^* \\ \sqrt{\kappa}\varepsilon_{in} \\ \sqrt{\kappa}\varepsilon_{in}^* \\ \zeta_b \\ \zeta_b^* \\ \zeta_c \\ \zeta_c^* \\ \zeta_d \\ \zeta_d^* \end{bmatrix}}_{n(t)} \quad (7)$$

$\Gamma_s = (\kappa + i\Delta_{sp}); \Gamma_i = (\kappa + i\Delta_{ip}); \Gamma_b = (\gamma_s + i\Delta_s); \Gamma_c = (\gamma_{as} + i\Delta_{as}); \Gamma_{ph} = (\gamma_{ph} + i\Delta_{ph}); \Pi_1 = -i(\kappa_s\alpha_s + \kappa_i\alpha_i); \Pi_2 = -i(\kappa_{as}\alpha_s + \kappa_{ai}\alpha_i);$

This equation, in essence, can be simply introduced as $du(t)/dt = A_{i,j} \times u(t) + n(t)$, which is solved as $u(t) = \exp(A_{i,j}t)u(0) + \int(\exp(A_{i,j}s).n(t-s))ds$, where $u(t)$ is a column matrix indicated the field operators and their Hermitian conjugates, and $A_{ij}$ is a 10×10 matrix. Moreover, $n(s)$ is the noise column matrix shown in Eq. 7.

Eventually, the time evolution of the input field modes (signal and idler), Stoke mode, Anti-stoke mode, and phonon are calculated. Now, this is the time to study the photon bunching and anti-bunching effects, squeezing effect, and the entanglement between various modes.

Table I. Constants used in this work [21, 22, 24]

| | |
|---|---|
| $r_0$ | 15 nm |
| $r_1$ | 25 nm |
| $r_2$ | 30 nm |
| $\lambda_s$ | 532 nm |
| $\lambda_i$ | 650 nm |
| $\gamma_{21}$ | $10^5$ 1/s |
| $\gamma_{23}$ | $10^6$ 1/s |
| $\gamma_{31}$ | $5\times10^7$ 1/s |
| $\chi^{(2)}$ | $10^{-12}$ m/v |
| $\omega_{ph}$ | $10^9$ 1/s |
| $\kappa$ | $10^{13}$ 1/s |
| $\kappa_{stoke}$ | $10^9$ 1/s |
| $\kappa_{a-stoke}$ | $10^8$ 1/s |
| $\kappa_{ph}$ | $10^2$ 1/s |
| $\kappa_s=\kappa_i$ | $2.5\times10^8$ 1/s |
| $\kappa_{as}=\kappa_{ai}$ | $0.5\times10^8$ 1/s |
| $P_c$ | 10 mw |

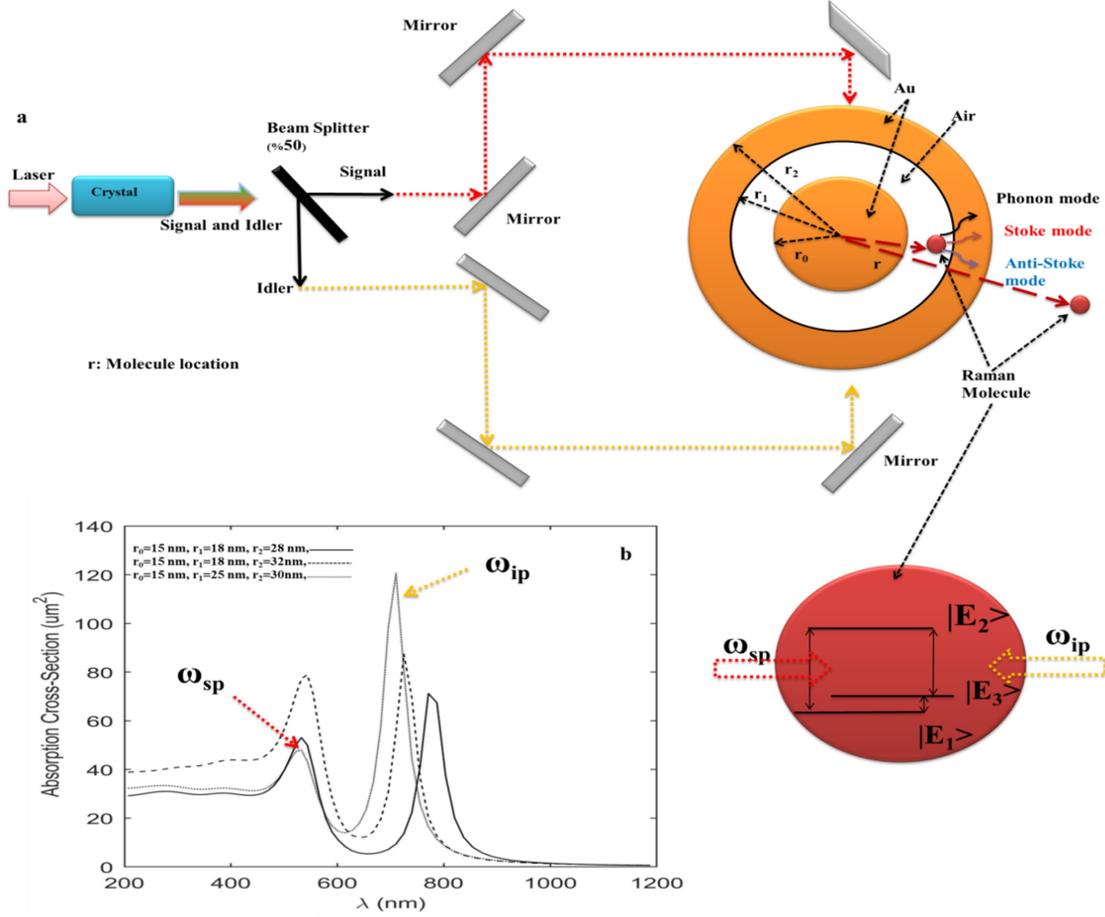

Fig. 1a) Schematic illustration of the interaction of the entangled photons with core/shell NPs and dye molecules incubated at the middle gap between two plasmonic materials; b) the core/shells NPs absorption cross-section.

*Squeezing effect*

A large number of theoretical and practical works have been considered to obtain the squeezed state [21]. With knowledge of this point, we want to study the Raman modes squeezing effect. Here, we assume that the possibility of squeezing of Raman modes is due to the entangled two-photon non-classicality coupled to the core/shell NPs and dye molecules. To investigate squeezing effect, we just need to consider the quadrature fluctuation for each field. Typically, we consider the quadrature operators for signal mode as:

$$X_{as} = \frac{1}{\sqrt{2}}[\delta a_s + \delta a_s^+], P_{as} = \frac{-i}{\sqrt{2}}[\delta a_s - \delta a_s^+] \quad (8)$$

where $\delta a_s$ and $\delta a_s^+$ are defined from Eq. 8 as:

$$\delta a_s = \sum_{j=1}^{10} A_{1,j}(t) u_j(0) + \sum_{j=1}^{10} \int_0^t A_{1,j}(s) n_j(s) ds,$$
$$\delta a_s^+ = \sum_{j=1}^{10} A_{2,j}(t) u_j(0) + \sum_{j=1}^{10} \int_0^t A_{2,j}(s) n_j(s) ds \quad (9)$$

In a similar way, one can calculate the quadrature operators for other modes such as idler, Stoke, Anti-stoke, and phonon modes. Here, the quadrature fluctuation is presented as $(\Delta X_{as})^2 = <X_{as}^2> - <X_{as}>^2$ and $(\Delta P_{as})^2 = <P_{as}^2> - <P_{as}>^2$. In the following, to calculate the quadrature fluctuation, signal mode for instance, we need to know about $<\delta a_s \delta a_s>$, $<\delta a_s \delta a_s^+>$, $<\delta a_s^+ \delta a_s>$, $<\delta a_s^+ \delta a_s^+>$, $<\delta a_s>$, and $<\delta a_s^+>$. Here, we just calculate $<\delta a_s(t) \delta a_s^+(t)>$, and others can be evaluated in a similar way:

$$<\delta a_s(t)\delta a_s^+(t)> = \sum_{j=1}^{10}\sum_{i=1}^{10} A_{1,j}(t)A_{2,i}(t)<u_j(0)u_i(0)> + \sum_{j=1}^{10}\sum_{i=1}^{10}\int_0^t A_{1,j}(s)A_{2,i}(s)<n_j(s)n_i(s)>ds$$

$$= A_{11}A_{22}<\delta a_s(0)\delta a_s^+(0)> + A_{13}A_{24}<\delta a_i(0)\delta a_i^+(0)> + A_{15}A_{26}<\delta b(0)\delta b^+(0)> + A_{17}A_{28}<\delta c(0)\delta c^+(0)> + A_{19}A_{210}<\delta d(0)\delta d^+(0)> \quad (10)$$

$$\int_0^t [A_{11}A_{22}\kappa<\varepsilon_{in}(s)\varepsilon_{in}^*(s)> + A_{13}A_{24}\kappa<\varepsilon_{in}(s)\varepsilon_{in}^*(s)> + A_{15}A_{26}<\zeta_b(s)\zeta_b^*(s)> + A_{17}A_{28}<\zeta_c(s)\zeta_c^*(s)> + A_{19}A_{210}<\zeta_d(s)\zeta_d^*(s)> +$$

$$A_{12}A_{21}\kappa<\varepsilon_{in}^*(s)\varepsilon_{in}(s)> + A_{14}A_{23}\kappa<\varepsilon_{in}^*(s)\varepsilon_{in}(s)> + A_{16}A_{25}<\zeta_b^*(s)\zeta_b(s)> + A_{18}A_{27}<\zeta_c^*(s)\zeta_c(s)> + A_{110}A_{29}<\zeta_d^*(s)\zeta_d(s)>]ds$$

In this equation, $<\delta a_s(0)\delta a_s^+(0)> = <\delta a_i(0)\delta a_i^+(0)> = <\delta b(0)\delta b^+(0)> = <\delta c(0)\delta c^+(0)> = <\delta d(0)\delta d^+(0)> = 1$. We have also introduced the optical and Raman mode's input noises which are given by $\varepsilon_{in}$, $\zeta_b$, $\zeta_c$, and $\zeta_d$, respectively. These input noises obeying the following correlation function [24].

$$<\varepsilon_{in}(s)\varepsilon_{in}^*(s')> = [N(\omega_{sporip})+1]\delta(s-s'); \quad <\varepsilon_{in}^*(s)\varepsilon_{in}(s')> = [N(\omega_{sporip})]\delta(s-s')$$

$$<\zeta_b(s)\zeta_b^*(s')> = \kappa_{stoke}[N(\omega_s)+1]\delta(s-s'); \quad <\zeta_b^*(s)\zeta_b(s')> = \kappa_{stoke}[N(\omega_s)]\delta(s-s')$$

$$<\zeta_c(s)\zeta_c^*(s')> = \kappa_{a-stoke}[N(\omega_{as})+1]\delta(s-s'); \quad <\zeta_c^*(s)\zeta_c(s')> = \kappa_{a-stoke}[N(\omega_{as})]\delta(s-s') \quad (11)$$

$$<\zeta_d(s)\zeta_d^*(s')> = \kappa_{ph}[N(\omega_{ph})+1]\delta(s-s'); \quad <\zeta_d^*(s)\zeta_d(s')> = \kappa_{ph}[N(\omega_{ph})]\delta(s-s')$$

In this equation, $N(\omega) = [\exp(\hbar\omega/k_BT)-1]^{-1}$; where $k_B$ and T stand for Boltzmann's constant and operation temperature, respectively [24]. Indeed, $N(\omega)$ is the equilibrium mean thermal photon numbers of the different modes. Moreover, $\kappa_{stoke}$, $\kappa_{a-stoke}$, and $\kappa_{ph}$ are the Stoke mode, Anti-stoke mode, and Phonon mode coupling to reservoir mode, respectively. Therefore, by replacing Eq. 11 into Eq. 10, we can calculate $<\delta a_s(t)\delta a_s^+(t)>$. Also, based on this approach, other cases for analyzing the quadrature fluctuation are attained. In the following, a few related simulation results for different modes are illustrated. It should be noted that all simulations and modeling in this work are based on some engineered and cited [21, 22, 24] information presented in Table I. The constants in Table I are selected in such a way that the system should be stable, meaning that all real part of eigenvalues of the system matrix should be negative. Fig. 2 illustrates the quadrature fluctuations for different modes (signal and stoke modes) at various temperatures. In this case, we suppose that the Raman molecules are embedded in the gap region between the plasmonic core and shell with r = 20 nm. From this figure, at the first glance, one can deduce that by increasing the temperature the squeezing quadrature fluctuation is dramatically decreased, suggesting that by increasing the temperature the amplitude of the applied noise is severely increased, leading to a drop in the degree of the non-classicality. Figs. 2e and 2f depict that in 100K, signal and stoke modes do not show any fluctuation. This behavior is contributed to the fact that in Eq. 7 the input noise amplitude is a dominant factor and severely affects the system response. More importantly, in the Stoke mode quadrature squeezing occurs, by which the fluctuation patterns demonstrate the regular revival and collapse. The revival, actually, is regularly repeated with time evaluation. It is known that the revival is a pure quantum phenomenon and cannot be demonstrated classically. The collapse and revival occurred in all Raman modes quadrature fluctuation may be due to the fact that the uncorrelated and correlated between the incident frequency and Raman modes frequency can arise. Moreover, we know that the Stoke and signal mode frequencies are different, so their revival and collapse pattern are various. Overall, we can briefly conclude that the non-classical properties of the entangled incident photon are efficiently coupled to the Raman modes (stoke mode in fig. 2) and eventually Raman modes show the non-classical properties.

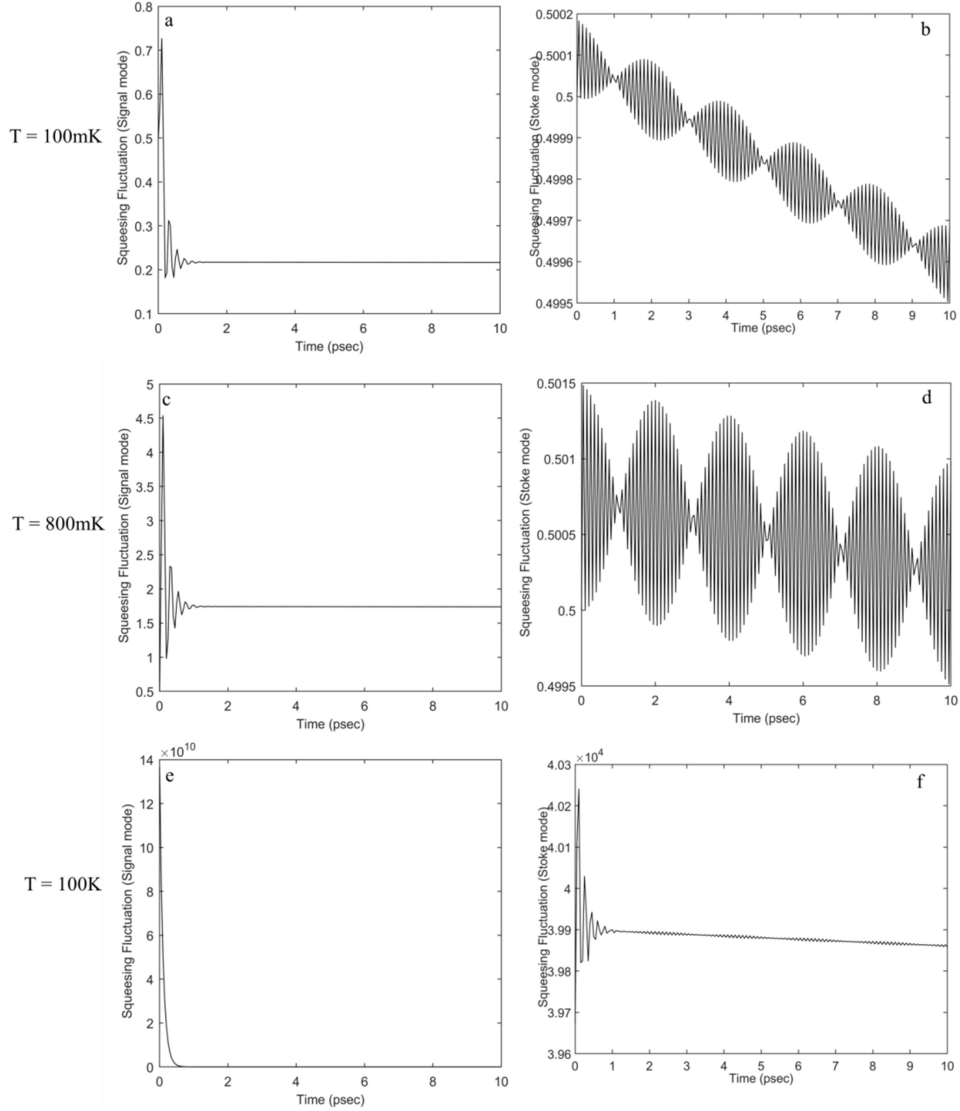

Fig. 2 Quadrature fluctuation for different temperature: (a), (c), and (e) Signal modes; (b), (d), and (f) Stoke mode; r = 20 nm.

*Photon bunching and anti-bunching effect:*
To study the non-classical properties of the system's modes, one can calculate the second-order correlation function at zero-time delay $g^{(2)}(\tau = 0)$ as:

$$g^{(2)}(\tau = 0) = 1 + \frac{<N^2> - <N>^2 - <N>}{<N>^2} \quad (12)$$

where $<N>$ is the average number of the photons in the considered field. If $g^{(2)}(\tau = 0)$ tends to unity, the corresponding photon distribution is Poissonian and hence the photon is un-bunched. However, the anti-bunching occurs just in the case of $g^{(2)}(\tau = 0) -1 < 0$, which means that the photon distribution is sub-Poissonian. Moreover, the non-classicality can occur if the second-order correlation function gets a negative quantity. We examine $g^{(2)}(\tau = 0)$ for Raman modes and input radiation field. Similar to the previous stage, for example, we calculate the second-order correlation function here just for signal mode as:

$$g^{(2)}{}_s(\tau = 0) = 1 + \frac{<N_s^2> - <N_s>^2 - <N_s>}{<N_s>^2} \quad (13)$$

where $<N_s> = <\delta a_s^+ \delta a_s>$ and $<N_s^2> = <\delta a_s^{+2}\delta a_s^2>$, which can be easily derived similar to Eq. 10. In the following, some of the modeling results for the second-order correlation function are illustrated (Fig. 3). Actually, in this figure, we want to study the bunching and anti-bunching of some modes in the contributed system, as well as the effect of the temperature. Similarly, we assumed that the Raman molecules are placed in the gap regions in the core/shell NPs. In Figs. 3a, 3b, 3c, and 3d, it is shown that when the environment temperature is small enough, all contributed modes show the anti-bunching effect, which is attributed to the non-classical properties of these modes. A comparison of Figs. 3b and 3d with Figs. 2b and 2d show that the anti-bunching coincides exactly at times with the occurrence of revival. This result can be explained by the correlation between modes, which happens at certain times. Furthermore, the anti-bunching pattern for the signal mode is different from that of the Anti-stoke modes. However, similar to the quadrature squeezing fluctuation behavior, when the temperature is increased, the considered modes show the bunching rather than anti-bunching. Hence, it can be stated that by increasing the environment temperature the second order correlation function goes to behave as a classical one.

*Entanglement between field modes:*
In this section, we study the entanglement between radiation fields and the Raman modes' separability. To measure two-photon mode entanglement, which is normally done by the Peres-Horodecki criterion [18-20], one can define the corresponding correlation matrix elements as:

$$V_{ij} = \frac{1}{2} <\hat{Y}_i\hat{Y}_j + \hat{Y}_j\hat{Y}_i> - <\hat{Y}_i><\hat{Y}_j>,$$
$$Y=[\hat{x}_1,\hat{p}_1,\hat{x}_2,\hat{p}_2], \hat{x}_i = \frac{1}{2}(a_i + a_i^+),$$
$$\hat{p}_{i\_out} = \frac{-i}{2}(a_i - a_i^+) \qquad (14)$$

In this equation, $V_{4\times4} = [A, C; C^T, D]$; where A, B, and C are 2×2 matrixes using which the two-output modes separability measurement is analyzed. All of the correlation matrix elements can be easily constructed using Eq. 10. To analyze the bipartite Gaussian system separability, we use the Symplectic eigenvalue, which is given by [25]:

$$\eta = \frac{1}{\sqrt{2}}\sqrt{\sigma(v) \pm \sqrt{\sigma(v)^2 - 2\det(v)}}, \qquad (15)$$
$$\sigma(v) = \det(A) + \det(B) - 2\det(C)$$

This equation is an important criterion for identification of the entanglement between two modes. Using this equation, we found that if $2\eta > 1$ the considered modes are purely separable; otherwise for $2\eta < 1$ two modes are entangled [25]. Therefore, one can find that which of two modes are entangled, and which ones are not. The simulated and modeled results are presented in Fig. 4. In the following, we study the entanglement between different modes in this system such as "signal and idler", "signal and phonon", and "stoke and anti-stoke", and also the effect of the environment temperature and Raman molecules location. From Fig. 4, which examines the effect of the temperature, it is shown that the "signal and idler" modes after interaction with NPs in the system remain entangled if the temperature is to remain small enough. Similarly, this can also occur for "signal and phonon" and "stoke and anti-stoke" modes. As can be seen in Figs. 4a-4f, when the temperature is increased from 100 mK to 800 mK, all contributed modes in the system remained entangled, suggesting that $2\eta$ should be smaller than unity. Moreover, in Figs. 4c and 4f, similar to stoke and anti-stoke quadrature squeezing and anti-bunching behaviors in Fig. 2 (b, d) and Fig. 3 (b, d), the revival and collapse effect are presented in a regular way. It is, however, noticeable that by increasing the temperature, the width of the revival phenomenon in "stoke and anti-stoke" entanglements is reduced; and this factor will definitely vanish when the temperature is grown enough. Furthermore, for the entanglement between "signal and idler", by increasing the temperature for a small time duration, two modes become separable while the alteration of the temperature does not affect in the case of "signal and phonon" entanglement. However, in Figs. 4g, 4h, and 4k, when the temperature rises up to a high level, $2\eta$ is dramatically increased for all modes. Additionally, the revival and collapse phenomenon in "Stoke and Anti-stoke" modes is destroyed, suggesting that the contributed modes become completely classical.
In the following, the effect of the Raman molecules location is investigated (Fig. 5). Clearly, by changing the Raman molecules location, the NPs-molecules coupling factor will be changed, implying that it incidentally alters the system's dynamics. In Fig. 5, the first row is for r = 16 nm, where the Raman molecules location is so close to the core, and the second case is for r = 24 nm, where the contributed location for the molecule is approximately near to the inner shell. By a comparison between the first and second rows, one can find that the alteration of the NPs-molecules coupling can slightly change the entanglement degree. This result can be explained by the fact that the plasmonic intensity is so high near the core region [7, 8] and, therefore, the maximum coupling occurs between incident signal and the plasmonic mode. By considering the third row, in which the molecules location is out of NPs r = 50 nm, it is seen that the entanglement between "signal and idler" and also for "stoke and anti-stoke" is decreased. Hence, the NPs plasmonic coupling to Raman molecules has an astonishing role in entanglement between incident modes and Raman modes. Surprisingly, it should be noted that the entanglements between "signal and phonon" modes are unaffected. Hence, the modeling results purely answer the article's original question that how and by which method we can attain the entangled Raman modes.

**Conclusions**

In this work, the quantum features of the enhanced Raman system were analyzed. For this purpose, we defined a system in which core/shell NPs containing the Raman molecules were excited by the entangled two-photon wave. By this mean, the incident wave initially is coupled to the NPs plasmonic field and then the plasmonic entangled mode is coupled to Raman molecules. The original aim of the present study was to investigate the Raman modes non-classicality and their entanglement. To this end, some quantum features such as quadrature squeezing, anti-bunching, and entanglement between modes were analyzed. It is surprisingly concluded that the incident mode non-classicality could be coupled to Raman modes, through which Raman modes demonstrated non-classical properties such as quadrature fluctuation and anti-bunching effect. Moreover, some of the Raman modes such as Stoke and Anti-stoke displayed a revival behavior, which is one of the quantum phenomena. In these cases, by the passage of time, the revival rose regularly maybe due to the Raman modes correlated with the incident mode.

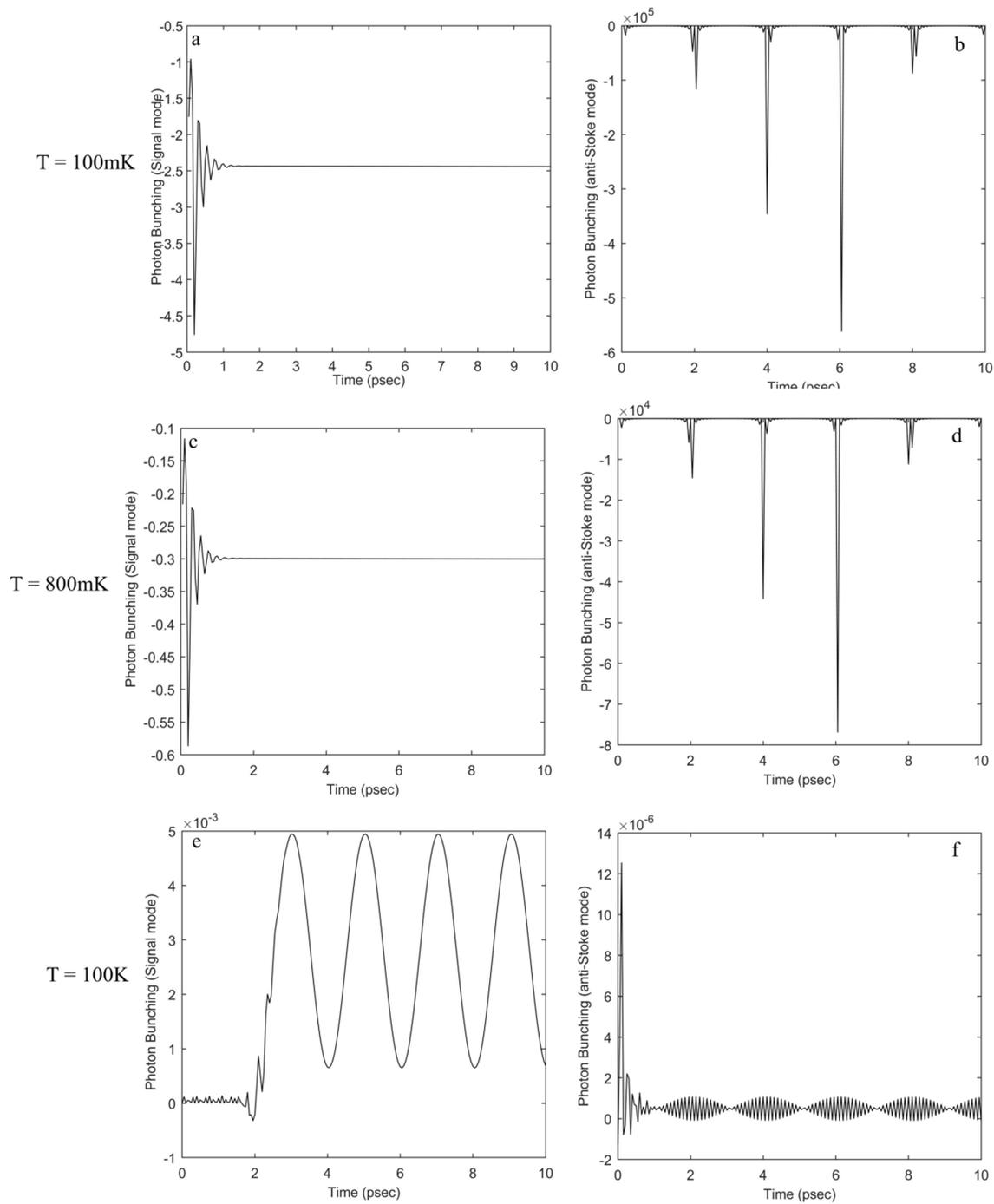

Fig. 3 Photon bunching and anti-bunching for different temperature: (a), (c), and (e) Signal modes; (b), (d), and (f) Stoke mode; r = 20 nm.

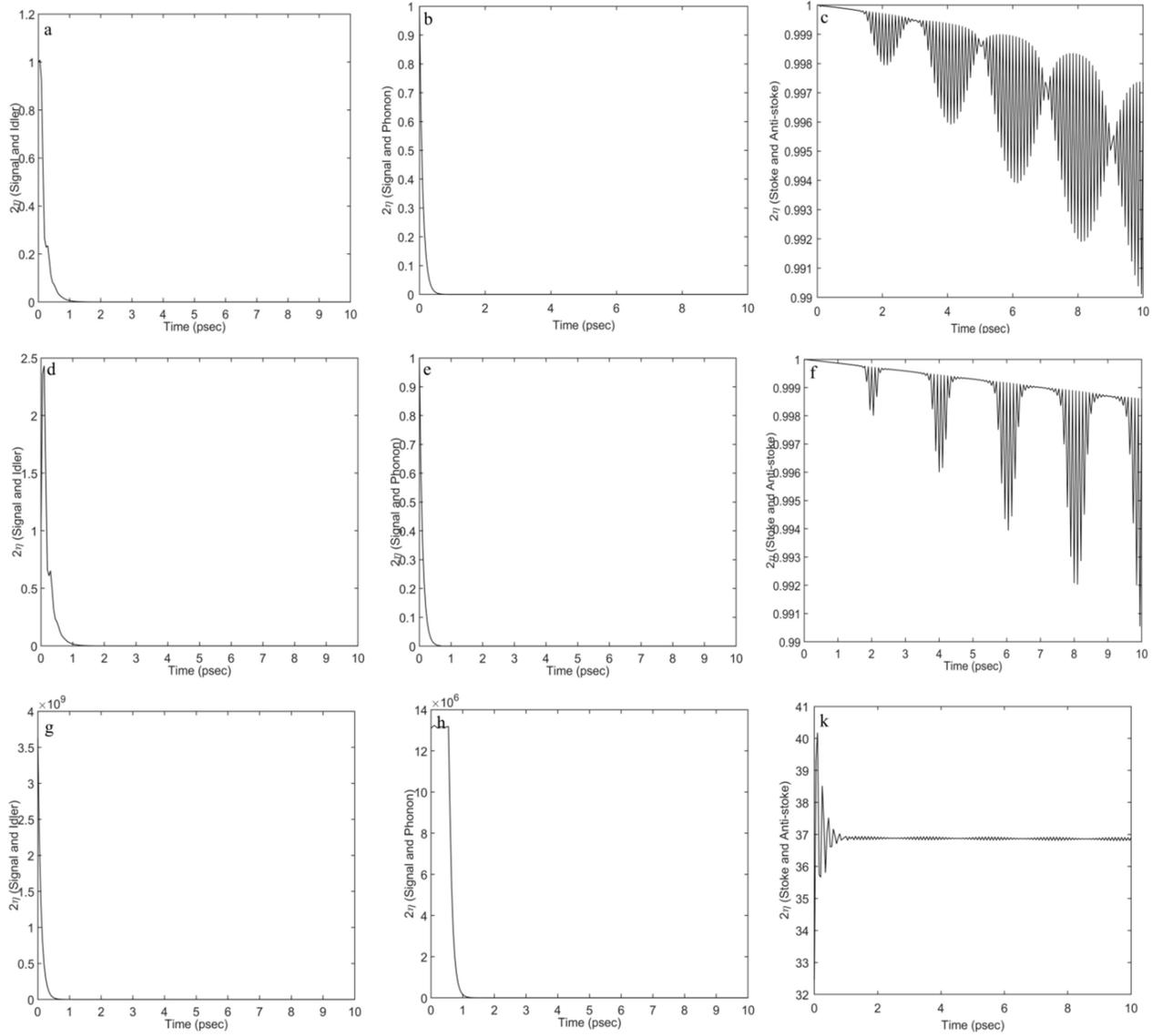

Fig. 4 Temperature effect on Entanglement for various modes (First row: 100 mK, Second row: 800 mK, Third row: 10 K): (a), (d), and (g) signal and idler modes; (b), (e), and (h) signal and phonon modes; (c), (f), and (k) stoke and anti-stoke modes; r = 20 nm.

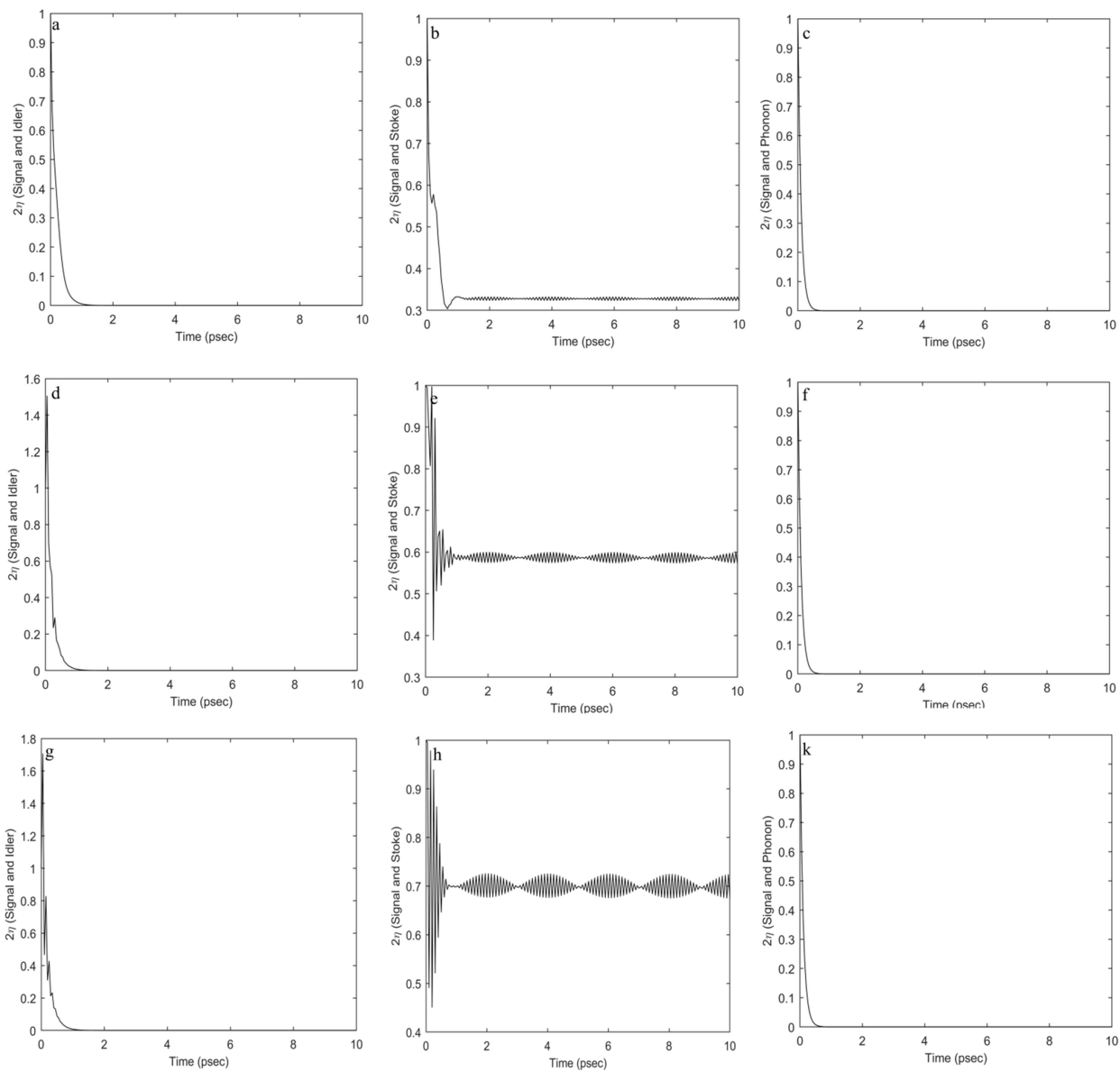

Fig. 5 Raman molecules location effect on Entanglement for various modes (First row: 16 nm, Second row: 24 nm, Third row: 50 nm): (a), (d), and (g) signal and idler modes; (b), (e), and (h) signal and phonon modes; (c), (f), and (k) stoke and anti-stoke modes; Tem = 200 mK.

**Acknowledgement:** The author thanks Prof. Erhan Piskin due to his help and guidance.